\documentclass[aps,prd,showpacs,floatfix,twocolumn]{revtex4}

\usepackage{bm}
\usepackage{latexsym}
\usepackage{dcolumn}
\usepackage{amsmath,amsfonts,amssymb}
\usepackage{graphicx,epsfig}
\usepackage{verbatim}
\usepackage{color}
\def\nn{\nonumber}

\def\l{\left}
\def\r{\right}
\def\DM{\mathrm{d}}

\def \Rtt {\ {}^{(2)}R}
\reversemarginpar
\begin{document}
\title{Holographic repulsion and confinement in gauge theory}
\author{Viqar Husain$^{1,2}$}
\email{ vhusain@unb.ca}
\author{Dawood Kothawala$^2$}
\email{dkothawa@unb.ca}
 
\affiliation{$^1$ Perimeter Institute for Theoretical Physics, Waterloo, ON, Canada} 
\affiliation{$^2$ Department of Mathematics and Statistics, University
of New Brunswick, Fredericton, NB, Canada E3B 5A3} \pacs{...}

\begin{abstract}

We show that for asymptotically anti-deSitter backgrounds with negative energy, such as the AdS soliton and regulated negative mass AdS-Schwarzshild metrics, the Wilson loop expectation value in the AdS/CFT conjecture exhibits a Coulomb to confinement transition. We also show that the quark-antiquark ($q \bar q$) potential  can be interpreted as affine time along null geodesics on the minimal string world sheet, and that its  intrinsic curvature provides a signature of transition to confinement phase. The result demonstrates a UV/IR relation in that the boundary separation of the  $q \bar{q}$ pair exhibits an inverse relationship with the radial descent of  the world sheet into the bulk. Our results suggest a generic (holographic) relationship between confinement in gauge theory and repulsive gravity, which in turn  is connected with singularity avoidance in quantum gravity.  

\end{abstract}
\date{\today}
\maketitle
\vskip 0.5 in
\noindent
\maketitle
\section{Introduction}

Among the intriguing insights to come from studies in quantum gravity are the  so called gauge-gravity dualities. These  provide a map  between a gauge theory without  gravity in $D$ dimensions and  a theory with gravity in $(D+1)$ dimensions. Such dualities naturally incorporate the appealing notion of holography, a  direct realization of which is the AdS/CFT 
conjecture given by Maldacena \cite{ads-cft}.   This conjecture proposes a duality between a specific type of string theory on a $D+1$ dimensional AdS background, and a conformal field theory living on the timelike boundary of this spacetime.

 In essence, the diffeomorphisms preserving the asymptotic AdS structure of the background   generate conformal diffeomorphisms of the boundary, and it is this isomorphism between the symmetry group of AdS in $(D+1)$ dimensions and conformal group of $D$ dimensional flat spacetime which makes it possible to postulate an equivalence between the  corresponding theories at the quantum level. Since its proposal by Maldacena and its subsequent quantitative realization \cite{ads-hol-witten, gubser, adscft-review}, the AdS/CFT conjecture has been used to study a wide range of boundary quantum systems using only  the geometric tools of general relativity. More recently this has been extended to the study of condensed matter systems using gravitational theories. 
 
 In this paper we study one specific application of this duality, namely the prescription for computing the expectation value of the Wilson loop of the boundary gauge theory via  a gravity calculation  \cite{m-loop}. This gives the quark anti-quark interaction potential of the gauge theory.  The precise correspondence is
\begin{eqnarray}
\langle W_{\gamma} \rangle = \int D s_{\gamma} \exp{\l( -S_{\rm NG}\{s_{\gamma}\} \r)} 
\end{eqnarray}
where $\gamma$ is the boundary loop, $s_\gamma$ is a bulk world sheet with $\gamma$ as its boundary, and $S_{\rm NG}\{s_{\gamma}\}$ is the Nambu-Goto string action. The r.h.s of this equation is typically approximated at the saddle point, giving
\begin{eqnarray}
\langle W_{\gamma} \rangle = \exp{\l[ - \Biggl( S_{\rm NG}\{s_{\gamma (\rm min)}\} - S_{\rm reg} \Biggl) \r]}
\end{eqnarray}
where $s_{\gamma (\rm min)}$ is the minimal worldsheet area with the loop $\gamma$ as its boundary. This is  obtained by extremizing the classical string action $S_{\rm NG}$. $S_{\rm reg}$ in this expression is the action of  two rectangular worldsheets extending into the bulk that  have the two timelike edges of the Wilson loop as their boundaries. 
This subtraction corresponds to the energy of the two free quarks moving on the boundary. (The suffix ``reg" signifies the fact that in an AAdS spacetime, this term removes the divergence coming from the first term.)

This approach to computing a Wilson loop  provides another connection between classical AAdS solutions of a gravitational theory and a quantum  expectation value in  gauge theory. Of particular interest is the confining phase of gauge theory, in which the quark-antiquark potential is linearly proportional to their separation.  
There are many papers in the literature that focus on confinement from AdS/CFT. The ones most relevant to the present
work are \cite{vh-wloop} where the negative mass case was studied, and \cite{gutperle-et.al} where AdS-scalar field 
metrics were studied. A class of conditions on metric functions that give confinement were derived in \cite{kinar-sonnen-et.al} and some examples appeared in \cite{sonnen-wline}. In addition to these works, there are many others that focus on this
aspect of the AdS/CFT correspondence; (see  eg. \cite{Brandhuber:1998,Rey:1998,Drukker:1999}). In particular \cite{Klebanov:2000, Maldacena:2000} provide examples of supergravity geometries that give confinement, and provide instances of the 
connection between negative bulk energy and confinement. 

The confinement phase in gauge theory, which occurs at large coupling and low energy (the opposite end of asymptotic freedom)  is also interesting from another perspective, known as the UV/IR duality. This  aspect of holography  states that the high energy regime in the bulk theory corresponds to the low energy regime in the boundary theory, and vice-versa. This  in turn suggests that the low energy confinement phase  of gauge theory is connected with the high energy quantum gravity regime in the bulk.  (This connection may arise for some semi-classical geometries that exhibit bulk repulsion \cite{Klebanov:2000,Maldacena:2000}. )

This last consequence of holography is our motivation for studying confinement in gauge theory using geometries that
violate  energy conditions, for it is widely believed that quantum gravity should provide a cure for curvature singularities, but doing so is concomitant with violating energy conditions. In other words, singularity avoidance requires that quantum gravity be
repulsive at short distances. 
 
Our aim is to investigate   bulk geometries that exhibit the Coulomb to confinement phase transition in the boundary gauge theory. We focus specifically on the roles played by the geometry parameters, negative bulk energy, and  geometrical properties of  the minimal surface.  In the next section we review the basic construction of  \cite{m-loop}. In Sec. III  we present evidence of  a relationship between confinement and repulsion.  Sec. IV gives   a new interpretation of the potential as the difference between bulk and boundary affine time separation of the $q\bar{q}$ pair, and  also gives a possible signature of confinement  via the Ricci scalar of the intrinsic world sheet geometry.  In the last  section we conjecture a {\it generic} 
relationship between singularity avoidance in quantum gravity and  confinement in gauge theory, based on the evidence that  both require negative bulk energy.

\section{Bulk geometry and the Wilson loop}

We summarize here the basics of the calculation \cite{m-loop}, focusing on the planar metric 
\begin{eqnarray}
\DM s^2 = \frac{r^2}{\ell^2} \Biggl[ -f(r) \DM t^2 + \DM x^2 + \DM y^2 + \DM z^2  \Biggl] \;+\; \frac{\ell^2}{r^2} \frac{ \DM r^2}{g(r)}
\end{eqnarray}
where  the functions $f(r)$ and $g(r)$ are such that the space-time is asymptotically AdS$_5$. (The non-planar case is  studied in \cite{witten-phase} where the AdS analog of the Hawking-Page transition is described; a planar analog of this for the AdS soliton appears in \cite{surya-et.al}.)

The procedure is to calculate  the Nambu-Goto action for a worldsheet which is bounded by a rectangular loop in a timelike plane on the boundary.  We take this loop to lie in the $t-x$ plane for definiteness. The sides of the rectangle are $T$ and $L$ along $t$ and $x$ respectively; $x \in [-L/2, +L/2]$. This loop has the standard interpretation of describing interaction between a quark-anti quark pair in the limit $T \rightarrow \infty$. As described in the previous section, the expectation value of the Wilson loop, in Maldacena's prescription, is given by the path integral of Nambu-Goto action, and is dominated by the minimal surface in the classical limit.

In the conventional static gauge, the worldsheet is described by a single function $r(x)$, and the induced metric is 
\begin{eqnarray}
h_{\mu \nu} \DM x^{\mu} \DM x^{\nu}  = 
 - \frac{r^2 f(r)}{\ell^2} \DM t^2 + \l(  \frac{r^2}{\ell^2} + \frac{\ell^2}{r^2 g(r)} \l(\frac{\partial r}{\partial x}\r)^2  \r) \DM x^2\nn
 \\
\end{eqnarray}
Therefore, the Nambu-Goto action becomes 
\begin{eqnarray}
S_{NG} = T \int \DM x \sqrt{ \frac{r^4 f(r)}{\ell^4} + \frac{f(r)}{g(r)} r'^2 }
\end{eqnarray}
where $r'={\partial r}/{\partial x}$ and the integral over $t$ is trivial due to $t$ independence of the integrand. To find the minimal surface, we note that the integrand does not depend on $x$ explicitly, and hence the ``Hamiltonian" corresponding to $x$ translations is a constant. This constant can be fixed by evaluating the Hamiltonian at the minimal radius, $r_m$, of the worldsheet, where $r'=0$. This gives 
\begin{eqnarray}
r' = \frac{r^2}{\ell^2} \sqrt{ \frac{r^4}{r_m^4} \frac{f g}{f_m} - g}
\label{rprime}
\end{eqnarray}  
where $f_m=f(r_m)$.  
With the requirement that the  worldsheet ends on the boundary loop, we have the condition $x(\infty)=\pm L/2$. Also, from symmetry, the minimal surface is symmetric about $x$, hence $x(r_m)=0$. Therefore the last equation gives 
\begin{equation}
\int_0^{L/2} \DM x = \int_{r_m}^{\infty} \frac{\ell^2}{r^2} \frac{1}{\sqrt{ \frac{r^4}{r_m^4} \frac{f g}{f_m} - g}} \DM r
\end{equation}
which gives 
\begin{eqnarray}
L(r_m) = \frac{2 \ell^2}{r_m} \int_{1}^{\infty}  \frac{\DM y}{y^2} \frac{1}{\sqrt{g(y;r_m)}} \frac{1}{\sqrt{ \frac{f(y;r_m)}{f_m(r_m)} y^4  - 1}}
\end{eqnarray}
where $y=r/r_m$, and  $f_m(r_m)=f(1;r_m)$.  This equation gives the quark-antiquark separation $L$ as a function of the 
minimum radial coordinate value of the world sheet.

The interaction potential $V(r_m)$  is obtained using the prescription
\begin{equation}
S_{NG}=T V(L)
\end{equation}
with the action computed for the minimal surface. Using $r'(x)$ from Eqn. (\ref{rprime}) gives 
\begin{equation}
\frac{S_{\rm NG}}{T}  = 2 \int_{r_m}^{\infty} \frac{\DM r}{r'}  \sqrt{ \frac{r^4 f(r)}{\ell^4} + \frac{f(r)}{g(r)} r'^2 }
\end{equation} 
which after some simplification leads to
\begin{eqnarray}
S_{\rm NG} &=& \frac{2 Tr_m}{f_m(r_m)} \nn\\
&& \times \int_1^{\infty} \DM y    \frac{ y^2 f(y;r_m)}
{\sqrt{g(y;r_m)}    \sqrt{ \frac{f(y;r_m)}{f_m(r_m)} y^4 - 1 }}
\nn \\
\end{eqnarray} 
To obtain the interaction potential from this, we first need to regularize this expression by subtracting the contribution coming from the edges of the loop along the $t$ direction, which has the interpretation of energy of free quarks. Since these 2 worldsheets lie in the $t-r$ plane, the corresponding Nambu-Goto action is 
\begin{equation}
S_{\rm reg} = 2 T \int_{\Lambda_c}^\infty \DM r \sqrt{\frac{f}{g}}
= 2 T r_m \int_{\Lambda_c/r_m}^{\infty} \DM y \sqrt{\frac{f}{g}}
\end{equation}
where $r=\Lambda_c$ is some cut-off radius upto which these free-quark worldsheets extend in the bulk. In pure AdS bulk, one can take $\Lambda_c=0$, whereas for a black hole in AdS, $\Lambda_c=r_0$,  the horizon radius. We can now calculate the potential $V(r_m)$ by subtracting this contribution from the expression for minimal worldsheet contribution.
\begin{eqnarray}
V(r_m) &=& \frac{1}{T}\ (S_{\rm NG} - S_{\rm reg}) 
\end{eqnarray}
Putting everything together we have 
\begin{widetext}
\begin{eqnarray}
\frac{V(r_m)}{\hbar c r_0/\alpha'} &=& \frac{2 r_m}{r_0} 
\l[ \;\; 
\int \limits_1^{\infty} \DM y \sqrt{\frac{f(y;r_m)}{g(y;r_m)}} \l( \sqrt{\frac{f(y;r_m)}{f_m(r_m)}} \frac{y^2}{\sqrt{ \frac{f(y;r_m)}{f_m(r_m)} y^4 - 1 }}  - 1 \r) 
- 
\int \limits_{\Lambda_c/r_m}^1 \DM y \sqrt{\frac{f(y;r_m)}{g(y;r_m)}} 
\;\; 
\r]
\nonumber \\
\frac{L(r_m)}{\ell^2/r_0} &=& \frac{2 r_0}{r_m} \int_{1}^{\infty}  \frac{\DM y}{y^2} \frac{1}{\sqrt{g(y;r_m)}} \frac{1}{\sqrt{ \frac{f(y;r_m)}{f_m(r_m)} y^4  - 1}} 
\label{VLrm}
\end{eqnarray}
\end{widetext}
where we have put in all the constants such as string length scale, $\ell_s=\sqrt{\alpha'}$, $\hbar$ and $c$. These are the equations we focus on in the remainder of the paper. In all the plots below, we have plotted the dimensionless combinations on the left, as a function of $r_m/r_0$ (which is the only combination on which the RHS will depend, since the entire $\ell$ dependence has been moved to the LHS).

\section{Negative energy and confinement}

Since our main interest in this work is the Coulomb to confinement transition and its relation to negative energy in the bulk,
we first present  evidence for this, before attempting a geometric  understanding in the following  section.
We consider three cases that exhibit this, the negative mass Schwarzschild-AdS metric, a regulated singularity free version
of it, and the AdS soliton.   

A curious point noted in \cite{vh-wloop} is that  if $f$ and $g$ are chosen to correspond to the negative mass AdS-Schwarzschild solution, ie. $f(r) = 1 + (r_0/r)^4 = g(r)$, then $V(L)$  shows a transition from Coulomb to confinement phase. The result is intriguing since the bulk metric has a naked singularity, and one would not have expected any sensible result, let alone a confining phase, to come from such a pathological bulk geometry.  

A relevant question is whether this result, which appears in Figs. 2, is connected to either the presence of a naked singularity,  the absence of a horizon, or to the negative energy, or all three. (It is known that the AdS black hole solution  does not lead to a confinement phase -- the $V(L)$ curve becomes multiple valued with respect to $L$ and the result is interpreted as a
screened Coulomb potential.) 

To investigate the issue further, we note that the above expressions for $V$ and $L$  suggest that the qualitative features of $V(L)$   depend only on $f(r)$, whereas all dependence on $g(r)$ is separated out in smooth analytic form. Therefore, one may imagine that the curvature singularity can be removed by regulating the function $g( r)$, while maintaining the asymptotic form of the metric, and at the same time  ensuring that one does recover a confinement phase.  This turns out to be possible
for a wide range of functions $g$.  

However, a further restriction on the choice of functions arises upon noticing two things. First, in Eqs.~(\ref{VLrm}), the square root in the denominator depends on $y^4$, and second, that confinement is obtained for a negative mass AdS-Schwarzschild
solution. This means that this forth power  appears to be important for the confinement result. In fact it is possible to check this numerically. Based on these observations, we make the following ansatz to remove the singularity at $r=0$:
\begin{eqnarray}
f(r) &=& 1 + \frac{r_0^4}{r^4}
\\
g(r) &=& 1 + \frac{r_0^4}{r^4} \Biggl[ 1 - Q(r/\lambda)  \Biggl]
\end{eqnarray}
where the length  $\lambda$ is some regulator scale. The finiteness of Kretschmann scalar at $r=0$ is ensured by putting the conditions $Q(0)=1$ and $Q'(0)=Q''(0)=Q'''(0)=0$ on the otherwise arbitrary function $Q$. Further, to maintain the AAdS form of the metric, we must have $\lim \limits_{r \rightarrow \infty} r^{-4} Q(r/\lambda)=0$. The specific choice we shall make to exhibit the plots etc. is 
\begin{eqnarray}
f(r) &=& 1 + \frac{r_0^4}{r^4}
\\
g(r) &=& 1 + \frac{r_0^4}{r^4} \Biggl[ 1 - \exp{ \l(-  \frac{r^4} {\lambda^4} \r) } \Biggl]
\end{eqnarray}
The Kretschmann scalar is finite everywhere, in particular at $r=0$ the behaviour is 
\begin{eqnarray}
R_{abcd} R^{abcd} ~ \overset{r=0}{\longrightarrow} ~ \frac{40}{\ell^4} \l[ 1 + \l( \frac{r_0}{\lambda} \r)^4 \r]^2
\end{eqnarray}

In addition to this regulated negative mass solution, we will also consider the AdS soliton \cite{adssoliton}  with 
metric   
\begin{equation}
\DM s^2 = \frac{r^2}{\ell^2} \Biggl[ - \DM t^2 + \DM x^2 + f( r)\DM y^2 + \DM z^2  \Biggl] \;+\; \frac{\ell^2}{r^2} \frac{ \DM r^2}{g( r)}
\end{equation}
with $f = g = 1- r^4/r_0^4$.  This solution has a negative energy, which unlike the negative mass Schwarzschild case,
is bounded below. The  worldsheet we consider has the same boundary  loop in the $t-x$ plane, but  its induced
 metric is now 
\begin{equation}
\DM s^2 = -\frac{r^2}{\ell^2}  \DM t^2 + \left( -\frac{r^2}{\ell^2} \;+\;  r'(x)^2 \frac{\ell^2}{r^2 g( r)}\right)   \DM x^2
\end{equation}
Therefore the boundary potential  for the soliton is obtained by simply setting $f=1$ and $g=1-r_0^4/r^4$ in the formulas  (\ref{VLrm}). 

Let us first note that for sufficiently large $r_m$ the world sheet does not probe too deep into the AdS bulk, and so 
we expect that $V(L)$ behaviour in this limit should be that of the pure AdS case, i.e. Coulomb.  Therefore deviations from Coulomb behaviour
should come from world sheets that descend deeper into the bulk.  This suggests looking at  the integrands in the formulas
for  $L$ and $V$  for $r_m\ll 1$ and $\Lambda_c\ll1$.  Expanding the integrands in a Taylor series at $r_m=0$ for the
regulated negative mass Schwarzschild case  gives 
\begin{eqnarray}
V = \l( \frac{r_0}{\ell} \r)^2 L - \frac{2 \lambda^2 r_0^2}{\sqrt{ r_0^4 + \lambda^4 }} \frac{1}{\Lambda_c} + O(r_m^3, \Lambda_c^3)
\end{eqnarray}
This demonstrates the  confining behaviour, and gives the slope and intercept of the $V(L)$ line; the expansion that gives this result obviously cannot be done for the positive mass case due to the square roots in the integrands.  We note that the regulator and  cutoff lengths $\lambda$ and $\Lambda_c$ determine the intercept. 

Fig. 1 gives the plots of $V(r_m)$ and $L(r_m)$ for the negative mass AdS-Schwarzschild geometry. Since $r_m$ measures the descent of the minimal world sheet into the bulk,  it is apparent from this figure that the boundary scales $V$ and $L$ have an inverse relationship with $r_m$. This may be interpreted as an instance of the UV/IR relation, which states that high energy in the bulk corresponds to low energy on the boundary. 

Fig. 2  gives the potentials $V(L)$  for the negative mass AdS-Schwarzschild geometry and the AdS soliton.  It is apparent that the confining regions for large $L$ correspond to negative energy repulsion  in deep bulk region (ie. small $r_m$).  The termination of the AdS soliton line in Fig. 2 may be interpreted as a correspondence with the breaking of the streched
$q \bar{q}$ string; this does not happen for the negative mass Schwarzschild case because, unlike the AdS soliton, it has no  energy lower bound.   Another interesting feature is the significant difference in the intercepts of the confining
potentials. An  understanding of this in terms of the minimal world sheet geometry appears in Sec. III.  Lastly
Fig. 3 gives the $V(L)$ curve for the regulated negative mass case; this  is identical to the unregulated case for the 
range of $0.01 \le r_m \le 5$ in units of $r_0$ (which appear in Fig. 1).   

\begin{figure}[!htb]
\begin{center}
\scalebox{0.4}{\includegraphics{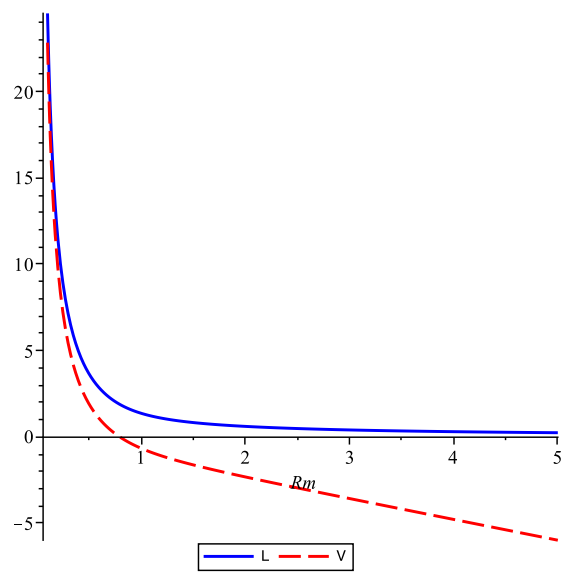}}
\label{Vrm}
\end{center}
\caption{$q \bar{q}$  potential $V(r_m)$ and their separation $L(r_m)$  for the
negative mass AdS-Schwarzschild geometry. This shows a UV/IR relation: large
$V$ and $L$ on the boundary correspond to small $r_m$ and vice versa.}
\end{figure}

\begin{figure}[!htb]
\begin{center}
\scalebox{0.4}{\includegraphics{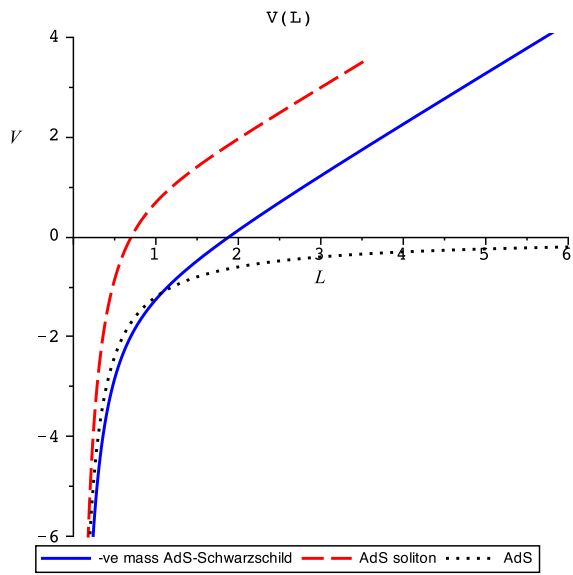}}
\end{center}
\caption{$q \bar{q}$    potential $V(L)$ from  AdS, AdS soliton and negative mass AdS-Schwarzschild geometries;
the latter two show a Coulomb to confinement transition. The AdS soliton line terminates at $r_m=r_0$, which corresponds to
its energy bound. }
\end{figure}

\begin{figure}[!htb]
\begin{center}
\scalebox{0.4}{\includegraphics{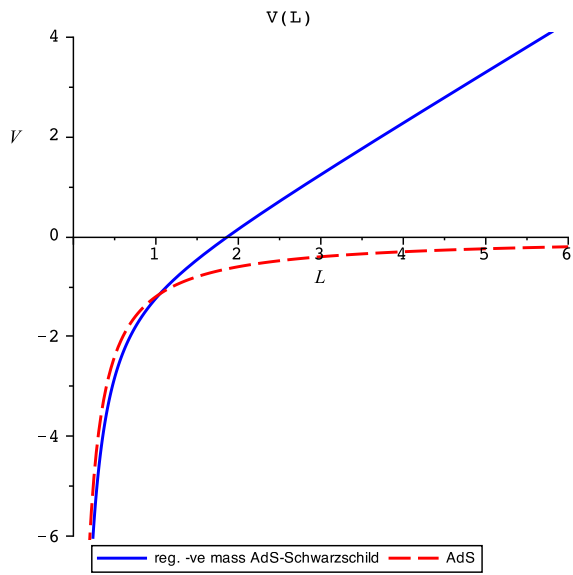}}
\end{center}
\caption{$q \bar{q}$   potential $V(L)$ for the regulated  negative mass AdS-Schwarzschild metric; it is identical
to the unregulated case for the range  $r_m$ plotted.}
\end{figure}

\noindent  {\it Numerical Fit:} It is   instructive to attempt a numerical fit to $V(L)$   to have some idea about order of magnitude of numerical factors involved. We give this below for the unregulated negative mass Schwarzschild case. The fit is of the form
\begin{equation}
2 \pi V_{\rm num}(L) = - a_0 - \frac{a_{-1}}{L} + a_{1} L
\end{equation}
where the constants are 
\begin{eqnarray}
a_0 &\approx& 0.9015 \l( \frac{\hbar c}{r_0} \r) \l(\frac{r_0}{\ell_s}\r)^2
\nn \\
a_{-1} &\approx& 1.2786 \  \hbar c \sqrt{2 g_{\rm YM}^2 N}
\nn \\
a_1 &\approx& 0.9547 \l( \frac{\hbar c}{r_0^2} \r) \frac{(r_0/\ell_s)^4}{\sqrt{2 g_{\rm YM}^2 N}}
\end{eqnarray}
where we have used the standard AdS-CFT dictionary to substitute $\ell^2 = \ell_s^2 \sqrt{2 g_{\rm YM}^2 N}$ in terms of $\ell_s$ and the gauge theory parameters $g_{\rm YM}$ and $N$. Note that the numerical factor $1.2786$ in the coefficient for $a_{-1}$ is very close to the pure AdS case for which $a_0=0=a_1$, and (numerical coefficient in $a_{-1}$) $\approx 1.4355$. Also, the length $L_c$ for which $V(L_c)=0$ is given by 
$L_c=(a_0/2 a_1) + \sqrt{(a_0/2 a_1)^2 + (a_{_{-1}}/a_1)}  \approx 1.72201 \l({2 g_{\rm YM}^2 N}\r)^{\frac{1}{2}} \ell_s^2/r_0$.

\section{Intrinsic geometry of the minimal worldsheet}

It is known that for $f=g=1$,  the parametric equation $V(r_m)$ and $L(r_m)$ lead exactly to the Coulomb phase
$V\sim 1/L$ \cite{m-loop}. Of much interest is the question of what $f$ and $g$ lead to confinement, i.e. $V\sim L$ and what the
transition region between the phases looks like. It is possible to  {\it algebraically} analyse the expressions of $V(r_m)$ and $L(r_m)$ to lay down conditions under which a given background will lead to Coulomb/confinement phase, and then verify what specific solutions fall into this class. This was done for example in \cite{kinar-sonnen-et.al}. However, the question we wish to consider 
is whether there is a {\it geometric} rather than algebraic way of characterizing this feature, since that might give a clearer insight into the whole issue. We show in this section that this is indeed possible using the intrinsic curvature and null geodesics of the minimal worldsheet.  

\subsection{$q \bar q$ potential and null geodesics}

We show that  $V(r_m)$ has a rather simple interpretation in terms of the affine parameter $\lambda$ along null geodesics $\bm k$ lying in the minimal worldheet. Consider the affine parameter such that $\bm k \cdot \bm\partial_t =-1$, i.e. normalized to have unit Killing energy with respect to the  ${\bm \partial}_t$ Killing vector of the bulk geometry. We shall see that  the potential is just the bulk affine time between the boundary quarks.  Interestingly,  the term that regulates the minimal world sheet action  has exactly the same interpretation. 

The induced geometry on the minimal and free quarks worldsheets can be expressed in the generic form  
\begin{eqnarray}
\DM s^2 &=& -N(n)^2 \DM t^2 + \DM n^2
\nn \\
\DM n &=& \Omega(r) \DM r
\label{eq:induced-met-gen-form}
\end{eqnarray}
and $\Omega(r)$ and $N(n(r))$ are known functions of $r$.   The null geodesics $\bm k$ for metrics of this form 
satisfying $\bm \nabla_{\bm k} \bm k=0$ are  given by 
\begin{eqnarray}
\bm k = C \l( \frac{1}{N^2} \bm \partial_t \pm \frac{1}{N} \bm \partial_n \r)
\end{eqnarray}
where the constant $C = - \bm k \bm \cdot \bm \partial_t$ is the Killing energy of these geodesics. We fix the scaling freedom in the choice of affine parameter by setting $C=1$, and  also choose (without any loss of generality due to $x \rightarrow -x$ symmetry) the $+$ sign above corresponding to outgoing geodesics. We therefore have 
\begin{eqnarray}
  N(n) \DM n = \DM \lambda
\end{eqnarray}

Let us denote  the interacting (minimal) and free quarks worldsheets respectively by $r_I(x)$ and $r_F(x)$. This gives the
following formulas for $N$ and $\Omega$. 

\begin{enumerate}

\item {\it Interacting quark worldsheet:} 

The minimal worldsheet is  given by Eqn. (\ref{rprime}) for which the induced metric is
\begin{eqnarray}
\DM s_I^2 &=& \frac{r_I^2 f(r_I)}{\ell^2} \l( - \DM t^2 + \frac{\ell^4}{ \l( f(r_I) r_I^4 - f_m r_m^4 \r)g(r_I)} \ \DM r_I^2 \r)
\nn \\
\label{eq:induced-geom-mws}
\end{eqnarray}
From this we have 
\begin{eqnarray}
N &\equiv& N_I\l(n(r_I)\r) = \frac{r_I} {\ell} \sqrt{f(r_I)}
\nn \\
\Omega &\equiv& \Omega_I(r_I) = \frac{\ell}{r_I} \l( \frac{r_I}{r_m} \r)^2 \sqrt{\frac{f(r_I)}{g(r_I)}} \times
\nn \\
&& \frac{1}{\sqrt{  (r_I/r_m)^4 f(r_I) - f_m }}
\end{eqnarray}

\item {\it Free quarks worldsheet:} The induced metric of the worldsheets corresponding to free quarks is  
\begin{eqnarray}
\DM s_F^2 &=& - \frac{r_F^2 f}{\ell^2} \DM t^2 + \frac{\ell^2}{r_F^2 g} \DM r_F^2
\end{eqnarray}
which gives
\begin{eqnarray}
N &\equiv& N_F(n(r_F)) = \frac{r_F \sqrt{f(r_F)}}{\ell}
\nn \\
\Omega &\equiv& \Omega_F(r_F) = \frac{\ell}{r_F \sqrt{g(r_F)}}
\end{eqnarray}

\end{enumerate}

We are now ready to establish the connection between the $q \bar q$ potential $V$ and the affine parameter $\lambda$ along the null geodesic $\bm k$.  Using the above expressions for $N_I, \Omega_I, N_F$ and $\Omega_F$,   $V$ may be written as 
\begin{eqnarray}
V &=& 2 \int \limits_{M}^{\mathcal B} \DM r_I \; \Omega_I N_I -  2\int \limits_{C}^{\mathcal B} \DM r_F \; \Omega_F N_F
\nn \\
&=& 2 \int \limits_{M}^{\mathcal B} \DM \lambda_I - 2\int \limits_{C}^{\mathcal B} \DM \lambda_F
\end{eqnarray}
where $\mathcal B, M$ and $C$ symbolically denote the values of the  integration parameter  at the boundary, minimal radius $r_m$, and the cut-off radius $\Lambda_c$ respectively. We have also suppressed the explicit dependence of r.h.s on 
$r_m$ for brevity. 

This is a  compact form of the expression for $V$, and comes with a nice geometric interpretation: the $q \bar q$ potential is (twice) the difference of affine distances along the null geodesics of the minimal worldsheet, propagating from minimal point to the boundary, and those on the free quark world sheet propagating from cut-off radius to the boundary. 

We note also that $L$ is (proportional to) the affine distance travelled by a null geodesic in the boundary, from $x=-L/2$ to $x=+L/2$. (In fact, if one rescales the metric near the boundary by $(\ell/r)^2$ and then take $r \rightarrow \infty$ limit, $L$ is exactly the affine distance since the boundary metric becomes flat.) Therefore, if $T_{sheet}$ and $t_{bound}$ denote the bulk and boundary affine times, then 
\begin{equation}
V(L) \equiv T_{sheet}(t_{bound})
\end{equation}  
This particular geometric interpretation of the $q \bar q$ potential might be relevant for interpreting the behaviour of the $V(L)$ curve in terms of a causal connection between events in the bulk and boundary.    

\subsection{Geometrical signature of confinement}

We now explore another feature of the minimal worldsheet geometry which appears to be a plausible signature of the  Coulomb to confinement phase.  The quantity we consider  is the intrinsic Ricci scalar $\Rtt$ of the minimal worldsheet  calculated at the radius $r_m$. For metrics of the form (\ref{eq:induced-met-gen-form}), we have
\begin{equation}
 \Rtt = -\frac{2}{N}\ \frac{\DM^2 N}{\DM n^2} 
 \end{equation}
For the minimal wordsheet given by (\ref{eq:induced-geom-mws}),  the value of $\Rtt$ at $y=r/r_m=1$ can be obtained
easily by noting that $\Omega_I^{-1} \rightarrow 0$ as $y \rightarrow 1$. This gives
\begin{eqnarray}
\Rtt_m &=& \Rtt|_{y=1} 
\nn \\
&=& - \frac{1}{\ell^2} \frac{g(1;\chi)}{f(1;\chi)} \l[ \frac{\DM}{\DM y} \ln N_I \r]_{y=1} \l[ \frac{\DM}{\DM y} \l(y^4 f\r) \r]_{y=1}
\nn \\
\label{eq:2Rlim}
\end{eqnarray}
where $\chi = r_m/r_0$ and $g(1;\chi) \equiv g(y;\chi)|_{y=1}$ etc.  

We examine  this quantity for the three bulk geometries studied in the last section, since as we have seen,
each  exhibits a Coulomb to confinement transition and each has negative energy. The formulas for $^{(2)}R$ are as follows:
 
\begin{enumerate}

\item {\it Negative mass AdS-Schwarzschild}

\begin{eqnarray}
\hspace{0.1in} \mathrm{unregulated}:  \hspace{0.2in} 
\Rtt_m &=& - \frac{4}{\ell^2} \l( \frac{\chi^4 - 1}{\chi^4 + 1} \r)
\nn \\
\hspace{0.1in} \mathrm{regulated}:  \hspace{0.2in} 
\Rtt_m &=& - \frac{4}{\ell^2} \l( \frac{\chi^4 - 1}{\chi^4 + 1} \r) \times
\nn \\
&& \l( 1 - \frac{\exp{\l[-(\frac{r_m}{\lambda})^4 \r] }}{\chi^4 + 1} \r)
\nn \\
\end{eqnarray}

\item {\it AdS Soliton}

\begin{eqnarray}
\Rtt_m &=& - \frac{4}{\ell^2} \l( \frac{\chi^4 - 1}{\chi^4} \r)
\end{eqnarray}

\item {\it Positive mass AdS-Schwarzschild}

\begin{eqnarray}
\Rtt_m &=& - \frac{4}{\ell^2} \l( \frac{\chi^4 + 1}{\chi^4 - 1} \r)
\end{eqnarray}

\end{enumerate}

\begin{figure}[!htb]
\begin{center}
\scalebox{0.75}{\includegraphics{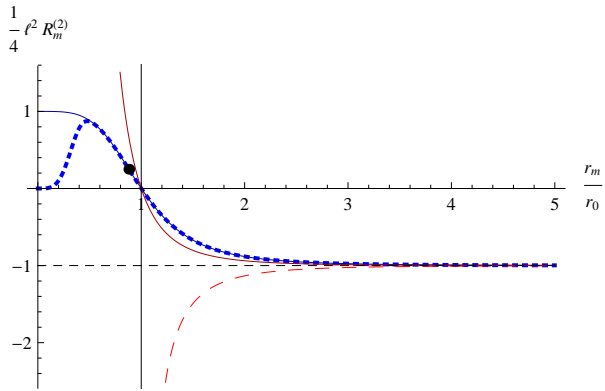}}
\label{fig:R}
\end{center}
\caption{Plot of $\Rtt$ at $r=r_m$ as a function of $r_m/r_0$.  (a) -ve mass AdS (blue) with the black dot representing inflection point of the curve (b) -ve mass regulated AdS (dotted (thick) blue) (c) +ve mass AdS (dashed red) (d) AdS soliton (dark red).  }
\end{figure}

The distinctive features  present  in these expressions and their graphs  are the following.
\begin{enumerate} 
\item[(i)] The first three of the above expressions suggest that, for solutions which do give a Coulomb to confinement transition, {\it $\Rtt$ at minimal radius changes sign at $r_m=r_0$}. 
\item[(ii)] As expected,   $r_m \gg r_0$  is the same for all four cases, ie. 
\begin{equation}
\Rtt_m \underset{r_m \gg r_0}{\longrightarrow} -4/\ell^2 
\nn
\end{equation}
Furthermore, for the negative mass AdS Schwarzschild case, $\Rtt_m$ goes from $-4/\ell^2$ for $r_m \gg r_0$ to $4/\ell^2$ for $r_m \ll r_0$, whereas for the AdS soliton it diverges as $r_m \rightarrow 0$.  This is however of no relevance since for the soliton, $r_m \in [r_0, \infty)$. 
\item[(iii)] For the positive mass AdS Schwarzschild solution, $\Rtt_m$ diverges on the horizon. This is  surprising
because all {\it bulk} invariants are known to be finite on the horizon surface. The divergence is perhaps a reflection of the fact that a {\it static} worldsheet configuration close to the horizon might require infinite stresses, just as an infinite acceleration is needed to hold a point particle near the horizon.
\item[(iv)] It is also curious that $\Rtt_m$ senses the presence of the scale $r_0$ even for the negative mass Schwarzschild solution, despite the fact that such a scale looks quite irrelevant from the point of view of bulk geometry (since the bulk metric depends only on $1+(r_0/r)^4$). Algebraically, it is easy to trace this result to the $(\ln N)$ term in Eq.~(\ref{eq:2Rlim}). 
\end{enumerate}

These cases suggest that a geometric signature of confinement is present in the intrinsic curvature $\Rtt_m$ of the minimal worldsheet at the minimal point; the distinguishing feature is the increase in this quantity as $r_m $ decreases, and, in particular, the change in sign (from $-$ve to $+$ve) at $r=r_0$. This in turn is reflection of the fact that the negative (quasi-local) energy region begins to be probed by the worldsheet. 
 
In fact, there is a more direct way of understanding the ``repulsion" associated with the negative energy solutions. To see this, note that, for the negative mass AdS black hole, $-g_{00}=(1/\ell^2)( r^2 + r_0^4/r^2 )$ has absolute minimum at $r=r_0$. Therefore, acceleration of a {\it static} particle, which is given by $a^i = (1/2) (g/f) \partial_r (-g_{00}) \delta^i_r$, changes sign from $+$ve to $-$ve as $r$ becomes less than $r_0$. That is, an {\it inward} radial force is needed to keep a particle static for $r<r_0$, implying repulsive gravity. In fact, the $(g/f) (\DM N/\DM y)$ term in expression (\ref{eq:2Rlim}) for $\Rtt_m$ is proportional to $a^r$, which provides a direct connection between this repulsive behavior and intrinsic worldsheet curvature, thereby strengthening our assertion that the latter captures the main feature (gravitational repulsion) responsible for transition to confinement. This remains true also for the AdS soliton ($g_{00}=-(r/\ell)^2; f=1$), in which case, although $g_{00}$ has no minima, $a^r$ changes sign at $r_0$ since $g(r)$ changes sign. The above argument also hints at an intriguing connection between quasi-local energy, effective potential governing point particle kinematics, and the signature of these features in the static worldsheet configuration.
 \\
 




\section{Discussion}

We have studied the $q\bar{q}$  potential modelled by  asymptotically AdS bulk geometries  with negative mass, both with
and without  curvature singularities. We have shown that $V(L)$ exhibits a transition from Coulomb to confinement phase, corresponding to the minimal string world sheet radius dipping from near the AdS boundary towards the $r=0$ region of the bulk. Therefore it appears that the low energy confinement phase corresponds to what would be considered the high energy sector of the bulk in the context of AdS/CFT duality. The essential points highlighted by the result presented here are: (i) confinement is a generic feature related to the presence of an additional  length scale (other than the AdS scale) in the bulk spacetime, which nevertheless does not correspond to a causally inaccessible region in the bulk with an event horizon. (ii) this result has nothing to do as such with the presence of naked singularities; indeed, as we have found  the singularity  can be regulated smoothly while retaining the seemingly important $r^4$ behaviour of the metric functions, without affecting the confinement phase, and (iii)  from the cases studied, it appears that  the key feature that gives rise to confinement is bulk negative energy repulsion. The latter is also associated with singularity avoidance in quantum gravity; ie. 
\begin{eqnarray}
&&\text{singularity avoidance in QG} \longleftrightarrow \text{negative energy}  \nn\\
&&\ \ \ \ \ \ \ \longleftrightarrow \text {confinement in gauge theory}. \nn
\end{eqnarray}
This plausible chain indicates that confinement in gauge theory might be intimately and generically connected with quantum gravity via holography, while  preserving  the UV/IR aspect of the AdS/CFT correspondence. 
 
\medskip

\noindent{\bf Acknowledgements} We thank Jaume Gomis and Rob Myers for  comments and discussion. This work was supported in part by the Natural Science  and Engineering Research Council of Canada. The work of D.K. was also supported by an AARMS fellowship. 

\bibliography{wloop-ref}

\end{document}